\documentclass{Interspeech_updated}

\usepackage{algorithm}
\usepackage{algorithmicx, algpseudocode}
\usepackage{cite}

\usepackage{lipsum}

\newcommand\blfootnote[1]{%
  \begingroup
  \renewcommand\thefootnote{}\footnote{#1}%
  \addtocounter{footnote}{-1}%
  \endgroup
}

\interspeechcameraready

\title{Few-step Adversarial Schr{\"o}dinger Bridge for Generative Speech Enhancement}

\author[affiliation={1}]{Seungu}{Han$^\dagger$}
\author[affiliation={1}]{Sungho}{Lee}
\author[affiliation={2}]{Juheon}{Lee}
\author[affiliation={1,3}]{Kyogu}{Lee}

\affiliation{Department of Intelligence
and Information}{Seoul National University}{Republic of Korea}
\affiliation{}{Supertone Inc.}{Republic of Korea}
\affiliation{Artificial Intelligence Institute}{Seoul National University}{Republic of Korea}
\email{hansw0326@snu.ac.kr, sh-lee@snu.ac.kr, juheon@supertone.ai, kglee@snu.ac.kr}

\usepackage{comment}

\begin{document}

\maketitle

\begin{abstract}
Deep generative models have recently been employed for speech enhancement to generate perceptually valid clean speech on large-scale datasets. Several diffusion models have been proposed, and more recently, a tractable Schrödinger Bridge has been introduced to transport between the clean and noisy speech distributions. However, these models often suffer from an iterative reverse process and require a large number of sampling steps---more than 50. Our investigation reveals that the performance of baseline models significantly degrades when the number of sampling steps is reduced, particularly under low-SNR conditions. We propose integrating Schr{\"o}dinger Bridge with GANs to effectively mitigate this issue, achieving high-quality outputs on full-band datasets while substantially reducing the required sampling steps. Experimental results demonstrate that our proposed model outperforms existing baselines, even with a single inference step, in both denoising and dereverberation tasks.
\end{abstract}
\blfootnote{$^\dagger$ Work partially done at Supertone Inc.}

\section{Introduction}
Deep generative models have been employed for various speech enhancement and restoration tasks~\cite{cdiffuse, universe, sgmse, se_sb, se_sb_sym, sgmse_sb}. This approach is motivated by the inherent nature of the problem: under severe conditions, such as low signal-to-noise ratio (SNR) scenarios, some information is lost, making it necessary to generate the missing parts~\cite{lowsnr_gen}.
Furthermore, unlike predictive models, which map inputs to a single deterministic output, generative models learn the underlying data distribution, allowing them to generate multiple valid estimates for a given input. This ability enhances their performance on non-intrusive metrics~\cite{nonintrusive}, making them valuable for real-world applications. Following the recent success in various domains, diffusion models or score-based generative models (SGMs)~\cite{sde}, have been proposed for speech enhancement~\cite{cdiffuse, universe, sgmse}. Diffusion models aim to learn the reverse process that transports the tractable prior distribution, typically Gaussian, into the data distribution. To better address the speech enhancement task, \cite{cdiffuse, sgmse} proposed conditional diffusion models, in which the reverse process begins with a mixture of noisy speech and Gaussian noise. However, these models suffer from a mismatch in the mean of the prior. 
To resolve this, the Schr{\"o}dinger Bridge (SB) framework~\cite{schrodinger, sb_sde, sb_forwardbackward, sb_matching}, whose goal is to find the path measure between one probability distribution and another, has been introduced as a natural solution. It enables the generative process to start directly from the noisy input~\cite{se_sb, se_sb_sym, sgmse_sb}. However, solving the general SB requires executing costly iterative procedures for simulating stochastic processes~\cite{sb_sde, sb_forwardbackward, sb_matching}, which limits its scalability and applicability.
On the other hand, in the special case where paired data is available, SB can be solved using a simulation-free approach~\cite{i2sb}. Several studies have built on this, exploring its incorporation into speech-related tasks, such as text-to-speech~\cite{sbtts} and audio super-resolution~\cite{sbsr}.
\looseness=-1
Furthermore, recent studies have demonstrated that SB outperforms diffusion models in speech enhancement for both wide-band~\cite{se_sb, se_sb_sym, sgmse_sb} and full-band~\cite{sbse_workshop} scenarios.

Diffusion models and the SB approach achieve high-quality results but suffer from slow inference speeds. 
This is due to the need for discretizing time into small step size to formulate each step of the reverse process as a Gaussian form.
On the other hand, \cite{se_sb} reported that the SB model is robust to the number of steps used in the reverse process for speech enhancement on wide-band data. However, our experiments reveal that this robustness does not always hold for full-band data, where a large number of iterations is still required, particularly in low-SNR cases.
Meanwhile, studies such as \cite{ddgan, siddm, ufogen} have explored combining Generative Adversarial Networks (GANs) with diffusion models, achieving both high-quality outputs and reduced sampling steps through adversarial training to match the denoising distribution.
Building on these advancements, we propose a novel approach that integrates the distribution matching GANs into the SB framework for speech enhancement. 
Our model not only avoids the prior mismatch, just like the original SB models, but also enables inference with as few as one to four steps. We evaluated our approach on full-band datasets, specifically using EARS-WHAM for denoising and EARS-Reverb for dereverberation~\cite{ears}.
Experimental results demonstrate that our model outperforms $50$-step SB model with only a single-step inference of our model. 
Notably, our model achieves a large margin of improvement over the SB baselines under low-SNR conditions. 
Audio samples are available at \url{https://bit.ly/4i06Pgm}.

\section{Background}

\subsection{Schr{\"o}dinger Bridge for Speech Enhancement} \label{sb_se}
The Schrödinger Bridge (SB) problem~\cite{schrodinger} considers the optimization of path measures with boundary conditions:
\begin{equation} \label{eq:sb_kl}
\min_{\mathcal{Q} \in \mathcal{P}_{[p_0,p_T]}} D_{\mathrm{KL}}(\mathcal{Q} || \mathcal{P}).
\end{equation}
Here, $\mathcal{P}$ denotes a reference measure, which is generally set to the path measure of the forward stochastic differential equation (SDE): $d\mathbf{x}_t = f(t, \mathbf{x}_t) dt + g(t) d\mathbf{w}_t, \mathbf{x}_0 \sim p_0$ as SGMs~\cite{sde}. The drift term is denoted by $f$, the diffusion term by $g$, and $\mathcal{P}_{[p_0, p_T]}$ is a set of path measures with marginal densities $p_0$ and $p_T$ at $t=0$ and $t=T$, respectively.
Then, the solution to the Eq.~\ref{eq:sb_kl} can be expressed by the path measure of a pair of forward and reverse SDEs~\cite{sb_forwardbackward}: 
\begin{subequations}
\thickmuskip=0.8\thickmuskip
\medmuskip=0.8\medmuskip
\begin{align}
    d\mathbf{x}_t &= [f(t, \mathbf{x}_t) + g^2(t) \nabla_{\mathbf{x}_t} \log \Psi(t, \mathbf{x}_t)]dt + g(t) \, d\mathbf{w}_t \\
    d\mathbf{x}_t &= [f(t, \mathbf{x}_t) - g^2(t) \nabla_{\mathbf{x}_t} \log \overline{\Psi}(t, \mathbf{x}_t)]dt + g(t) \, d\overline{\mathbf{w}}_t
\end{align}
\end{subequations}
where $\mathbf{x}_0 \sim p_0$ and $\mathbf{x}_T \sim p_T$.
While $f$ and $g$ are the same as in the reference SDEs, the additional nonlinear drift terms, $\nabla_{\mathbf{x}_t} \log \Psi(t, \mathbf{x}_t)$ and $\nabla_{\mathbf{x}_t} \log \overline{\Psi}(t, \mathbf{x}_t)$, are introduced.
The functions $\Psi$ and $\overline{\Psi}$ are time-varying energy potentials governed by coupled partial differential equations (PDEs):
\begin{subequations}
\thickmuskip=0.7\thickmuskip
\medmuskip=0.7\medmuskip
\thinmuskip=0.7\thinmuskip
\begin{align}
    \frac{\partial \Psi}{\partial t} &= -\nabla_{\mathbf{x}} \Psi^\intercal f - \frac{1}{2} \mathrm{Tr} \left(g^2 \nabla_{\mathbf{x}}^2 \Psi\right), \\
    \frac{\partial \overline{\Psi}}{\partial{t}} &= -\nabla_{\mathbf{x}} \cdot \left(\overline{\Psi} f\right) + \frac{1}{2} \mathrm{Tr} \left(g^2 \nabla_{\mathbf{x}}^2 \overline{\Psi}\right),
\end{align}
\end{subequations}
subject to the conditions $\Psi_0 \overline{\Psi}_0 = p_0$ and $\Psi_T \overline{\Psi}_T = p_T$.
The marginal of the SB state $\mathbf{x}_t$ is then given as $p_t=\Psi_t \overline{\Psi}_t$~\cite{sb_forwardbackward}. 
SB allows transportation between two arbitrary distributions beyond Gaussian priors. Thus, it is natural to formulate the speech enhancement task as an SB problem, where the goal is to transport between noisy and clean data distributions.

\looseness=-1
While the SB problem typically has no analytic solution, closed-form solutions exist for specific cases, such as when Gaussian boundary conditions are employed~\cite{i2sb, sbtts}. This formulation can be adapted to avoid the prior mismatch encountered in SGMs for speech enhancement. 
For that, \cite{se_sb, sgmse_sb} assume a linear drift $f(\mathbf{x}_t)=f(t) \mathbf{x}_t$ and Gaussian boundary conditions $p_0=\mathcal{N}(\mathbf{x};\mathbf{x}_0,\epsilon_0^2\mathbf{I})$ and $p_T=\mathcal{N}(\mathbf{x};\mathbf{y},\epsilon_T^2\mathbf{I})$ with $\epsilon_T=\exp(\smash{\int_0^Tf(\tau)d\tau})\cdot \epsilon_0$.
For $\epsilon_0 \to 0$, $\overline{\Psi}_t, \Psi_t$ converge to the tractable solution between clean data $\mathbf{x}_0$ and noisy data $\mathbf{y}$:
\begin{equation}
    \overline{\Psi}_t = \mathcal{N}(\alpha_t \mathbf{x}_0, \alpha_t^2 \sigma_t^2 \mathbf{I}), \quad  \Psi_t = \mathcal{N}(\bar{\alpha}_t \mathbf{y}, \alpha_t^2 \bar{\sigma}_t^2 \mathbf{I})
\end{equation}
where the parameters are given as  $\alpha_t = \exp \smash{(\int_0^t f(\tau)d\tau}), \sigma_t = \smash{\int_0^t {g^2(\tau)}/{\alpha_\tau^2} d\tau}, \bar{\alpha}_t=\alpha_t \alpha_T^{-1}$ and $\bar{\sigma}_t^2 = \sigma_T^2 - \sigma_t^2$.
Therefore, the marginal distributions of \(\mathbf{x}_t\) can be derived, and they follow Gaussian distributions, given as:  
\begin{equation} \label{eq:marginal}
q(\mathbf{x}_t|\mathbf{x}_0, \mathbf{y}) = \mathcal{N}(\mathbf{x}_t; w_\mathbf{x}(t)\mathbf{x}_0 + w_\mathbf{y}(t)\mathbf{y}, \sigma_x^2(t) \mathbf{I}),
\end{equation}
where the parameters are defined as 
\begin{equation} \label{eq:mu_sigma}
\sigma_x^2(t) = \frac{\alpha_t^2 \bar{\sigma}_t^2 \sigma_t^2}{\sigma_T^2}, \quad  w_\mathbf{x}(t) = \frac{\alpha_t \bar{\sigma}_t^2}{\sigma_T^2}, \quad  w_\mathbf{y}(t) = \frac{\bar{\alpha}_t \sigma_t^2}{\sigma_T^2}.
\end{equation}
Here, \(w_\mathbf{x}(t)\) and \(w_\mathbf{y}(t)\) are the weights for \(\mathbf{x}_0\) and \(y\), respectively, while \(\sigma_x^2(t)\) is the variance of the Gaussian distribution at time \(t\). Based on this design, the intermediate step can be sampled directly, and the same denoising objective as in the diffusion model can be applied. Moreover, the inference process starts with a noisy sample and iteratively follows the reverse SDE or the probability flow ordinary differential equation (ODE) to produce a clean speech estimate~\cite{se_sb, sgmse_sb}.

\subsection{Diffusion-GAN Hybrids}
Diffusion models require small step sizes, resulting in a large number of iterations to maintain the Gaussian assumption on the denoising distribution. In contrast, when the denoising step is large and the data distribution is non-Gaussian, the Gaussian assumption may not hold, resulting in a more complex distribution. To address this, DDGAN~\cite{ddgan} was the first to propose combining adversarial training with diffusion models, enabling denoising with larger discretization sizes and reducing the number of sampling steps. Continuous processes are discretized by selecting intermediate time points as $t_n=n/N$ with fixed $N$, $t_0=\epsilon$ and $t_N=T=1$. They leverage conditional GANs to optimize the divergence between the complex denoising distribution $q(\mathbf{x}_{t_{n-1}}|\mathbf{x}_{t_n})$ with $p_\theta(\mathbf{x}'_{t_{n-1}}|\mathbf{x}_{t_n})$:
\begin{equation} \label{eq:ddgan}
\min_{\theta} \mathbb{E}_{q(\mathbf{x}_{t_n})} [D_{\mathrm{adv}}(q(\mathbf{x}_{t_{n-1}}|\mathbf{x}_{t_n})||p_\theta(\mathbf{x}'_{t_{n-1}}|\mathbf{x}_{t_n}))].
\end{equation}

On the other hand, SIDDM~\cite{siddm} has been introduced to address scalability issues observed with DDGAN on large-scale datasets. 
The objective in Eq.~\ref{eq:ddgan} can be reformulated as matching the joint distributions $q(\mathbf{x}_{t_{n-1}}, \mathbf{x}_{t_n})$ and $p_\theta(\mathbf{x}'_{t_{n-1}}, \mathbf{x}_{t_n})$, since DDGAN indirectly aligns the conditional distributions $q(\mathbf{x}_{t_{n-1}} | \mathbf{x}_{t_n})$ and $p_\theta(\mathbf{x}'_{t_{n-1}} | \mathbf{x}_{t_n})$ through simple concatenation. Furthermore, the joint distributions can be factorized in the reverse direction as $q(\mathbf{x}_{t_{n-1}}, \mathbf{x}_{t_n}) = q(\mathbf{x}_{t_n} | \mathbf{x}_{t_{n-1}})\,q(\mathbf{x}_{t_{n-1}})$ and $p_{\theta}(\mathbf{x}'_{t_{n-1}}, \mathbf{x}_{t_n}) = p_{\theta}(\mathbf{x}_{t_n} | \mathbf{x}'_{t_{n-1}})\,p_{\theta}(\mathbf{x}'_{t_{n-1}}).$
Thus, joint distribution matching can be formulated as a combination of matching marginal distributions using adversarial divergence and matching conditional distributions using KL divergence:  
\begin{equation}  \label{eq:siddm}
\begin{aligned}
\min_{\theta} &\mathbb{E}_{q(\mathbf{x}_{t_n})} [D_{\mathrm{adv}}(q(\mathbf{x}_{t_{n-1}}) | p_{\theta}(\mathbf{x}'_{t_{n-1}})) \\
&+ \lambda_{\mathrm{KL}} D_\mathrm{KL}(p_{\theta}(\mathbf{x}_{t_n}|\mathbf{x}'_{t_{n-1}}) || q(\mathbf{x}_{t_n}|\mathbf{x}_{t_{n-1}}))].
\end{aligned}
\end{equation}
This factorization enables improved performance on complex data distributions.
Unlike the loss of DDGAN in Eq.~\ref{eq:ddgan}, where the discriminator incorporates $\mathbf{x}_{t_n}$, this approach excludes $\mathbf{x}_{t_n}$ for the discriminator’s input.

In Diffusion-GAN hybrids, the generator produces a sample of $\mathbf{x}_{t_{n-1}}$. Instead of directly producing $\mathbf{x}_{t_{n-1}}$ from the generator $G_\theta$, DDGAN and SIDDM parameterize $p_\theta(\mathbf{x}'_{t_{n-1}} | \mathbf{x}_{t_n})$ with posterior sampling of $q(\mathbf{x}_{t_{n-1}} | \mathbf{x}_{t_n}, \mathbf{x}'_0 = G_\theta(\mathbf{x}_{t_n}, t_n))$. Moreover, a modified parameterization is proposed by UFOGen~\cite{ufogen}, where \( p_\theta(\mathbf{x}'_{t_{n-1}}) = q(\mathbf{x}_{t_{n-1}} | \mathbf{x}'_0 = G_\theta(\mathbf{x}_{t_n}, t_n)) \), based on the forward process instead of the posterior. Starting with the objective of SIDDM in Eq.~\ref{eq:siddm}, this parameterization enables the model to be optimized, incorporating a simple reconstruction loss at $\mathbf{x}_0$:
\begin{equation}
\thickmuskip=0.5\thickmuskip
\medmuskip=0.5\medmuskip
\begin{aligned}
\min_{\theta} \mathbb{E}_{q(\mathbf{x}_{t_n})} [D_{\mathrm{adv}}(q(\mathbf{x}_{t_{n-1}}) | p_{\theta}(\mathbf{x}'_{t_{n-1}})) + \lambda_\mathrm{KL} || \mathbf{x}_0' - \mathbf{x}_0 ||^2 ].
\end{aligned}
\end{equation}
They hypothesize that SIDDM struggles with single-step generation due to the noise variance when sampling $\mathbf{x}_{t-1}$ with $\mathbf{x}_0$, whereas UFOGen enables successful one-step sampling by directly matching $\mathbf{x}_0$ in the reconstruction term. Note that the final loss can also be interpreted as training a diffusion model with adversarial refinement.

\section{Methods}

\subsection{Adversarial Schr{\"o}dinger Bridge}
\looseness=-1
We combine the Schr{\"o}dinger Bridge (SB) for speech enhancement with GANs. Since the SB is defined by a pair of SDE, it inherently satisfies the Markov property. Thus, the discrete version is obtained by discretizing the continuous process and then adapting it into the Diffusion-GAN Hybrids, in the same way as the usual SGMs. 
Furthermore, since noisy data $\mathbf{y}$ can serve as a condition in the speech enhancement task, the reverse process can be formulated with this condition at each step. In this case, the true denoising distribution \( q(\mathbf{x}_{t_{n-1}} | \mathbf{x}_{t_n}, \mathbf{y}) \) is matched with \( p_\theta(\mathbf{x}'_{t_{n-1}} | \mathbf{x}_{t_n}, \mathbf{y}) \), where each distribution is conditioned on $\mathbf{y}$. UFOGen can be naturally extended to incorporate condition, ensuring that the noisy input is consistently integrated into the model for speech enhancement.
The conditional transition probability is also Gaussian and is expressed as:  
\begin{equation} \label{eq:forward}
q(\mathbf{x}_{t_n}|\mathbf{x}_{t_{n-1}}, \mathbf{y}) = \mathcal{N}\left(\mathbf{x}_{t_n}; \boldsymbol{\mu}_{t_n|t_{n-1}}, \sigma_{t_n|t_{n-1}}^2 \mathbf{I}\right),
\end{equation}  
where the mean \(\boldsymbol{\mu}_{t_n|t_{n-1}}\) and variance \(\sigma_{t_n|t_{n-1}}^2\) are given by:  
\begin{subequations}
\thickmuskip=0.3\thickmuskip
\medmuskip=0.3\medmuskip
\thinmuskip=0.3\thinmuskip
\begin{align}
    \boldsymbol{\mu}_{t_n|t_{n-1}} &= \frac{w_\mathbf{x}(t_n) \mathbf{x}_{t_{n-1}}}{w_\mathbf{x}(t_{n-1})} + \bigg( w_\mathbf{y}(t_n) - \frac{w_\mathbf{x}(t_n)w_\mathbf{y}(t_{n-1})}{w_\mathbf{x}(t_{n-1})} \bigg)\mathbf{y}, \\
    \sigma_{t_n|t_{n-1}}^2 &= \sigma_{t_n}^2 - \left(\frac{w_\mathbf{x}(t_n)}{w_\mathbf{x}(t_{n-1})}\right)^2 \sigma_{t_{n-1}}^2
\end{align}
\end{subequations}
\looseness=-1
based on the mean and variance defined at each step in Eq.~\ref{eq:mu_sigma}.
The training process of our model is outlined in Algorithm~\ref{alg:ufogen_training} following the UFOGen scheme.
The model includes a generator $G_\theta$ and a discriminator $D_\phi$, conditioned on noisy data $\mathbf{y}$, and trained within an adversarial framework.
First, at a given time $t_n$, sample $\mathbf{x}_{t_{n-1}}$ using the marginal distribution in Eq.~\ref{eq:marginal} and then sample $\mathbf{x}_{t_n}$ using the conditional transition probability in Eq.~\ref{eq:forward}. Then, the generator produces a denoised estimate $\mathbf{x}'_0$ and sample $\mathbf{x}'_{t_{n-1}}$ through $q(\mathbf{x}_{t_{n-1}}|\mathbf{x}'_0, \mathbf{y})$. The discriminator and generator are updated using an adversarial loss to match intermediate data at time $t_{n-1}$ and a reconstruction loss. Instead of relying solely on the $l_2$-norm, we incorporate an auxiliary loss for the reconstruction loss, as done in \cite{se_sb, sgmse_sb}. 
Note that this framework can be applied to other diffusion processes that can be defined as a Markov chain, such as the Ornstein-Uhlenbeck Variance Exploding (OUVE) process~\cite{sgmse}. While Adversarial Schr{\"o}dinger Bridge Matching (ASBM)~\cite{sbgan} requires an explicit procedure for solving the Schr{\"o}dinger Bridge problem, our model does not, as it relies on a simulation-free framework.

\begin{figure}[t]
\vspace{-0.4cm}
\centering
\begin{algorithm}[H]
\caption{UFOGen Training}
\label{alg:ufogen_training}
\begin{algorithmic}[1]
\Require Generator $G_\theta$, Discriminator $D_\phi$
\Repeat\\
    $\mathbf{x}_0 \sim q(\mathbf{x}_0), n \sim \mathrm{Uniform}(1, \ldots, N), t_n = n/N$.\\
    $\mathbf{x}_{t_{n-1}} \sim q(\mathbf{x}_{t_{n-1}}|\mathbf{x}_0, \mathbf{y})$, $\mathbf{x}_{t_n} \sim q(\mathbf{x}_{t_n}|\mathbf{x}_{t_{n-1}}, \mathbf{y})$.\\
    $\mathbf{x}'_{t_{n-1}} \sim q(\mathbf{x}_{t_{n-1}}|\mathbf{x}'_0, \mathbf{y})$ where $\mathbf{x}'_0 = G_\theta(\mathbf{x}_{t_n}, \mathbf{y}, t_n)$.\\
    Update $D_\phi$ with gradient:
    \vspace{-2mm}
    \begin{equation}
    \begin{split}
        \nabla_\phi\Bigl(-\log(D_\phi(\mathbf{x}_{t_{n-1}}, & \mathbf{y}, t_{n-1})) \\ - &\log(1 - D_\phi(\mathbf{x}'_{t_{n-1}}, \mathbf{y}, t_{n-1}))\Bigr). \notag
    \end{split} 
    \end{equation}  \\
    \vspace{-3mm}
    Update $G_\theta$ with gradient:
    \vspace{-2mm}
    $$
    \nabla_\theta\left(-\log(D_\phi(\mathbf{x}'_{t_{n-1}}, \mathbf{y}, t_{n-1})) + \lambda_\mathrm{recon} L_\mathrm{recon}\right).
    $$
\vspace{-0.6cm}
\Until{converged}
\end{algorithmic}
\end{algorithm}
\vspace{-0.95cm}
\end{figure}

\subsection{Implementation Details}
\looseness=-1
For the noise schedule, the Variance Exploding (VE) schedule is chosen, following \cite{se_sb, sgmse_sb}, given by $g^2(t) = c k^{2t}$ and $f(t) = 0$. This results in $\alpha_t = 1$ and $\sigma^2_t = c(k^{2t}-1) / 2 \log k$. The recommended hyperparameters from~\cite{se_sb} are employed for the parameterization of the diffusion process. We select $N=4$ for the discretization, as \cite{ddgan, siddm, ufogen} reported it to yield the best results. 

The generator is based on the Noise Conditional Score Network (NCSN++) architecture. It utilizes power-compressed complex spectrograms as the input representation and uses the same model hyperparameters and STFT parameterization as in \cite{ears}, consistent with the baseline models.

We employ a multi-scale short-time Fourier transform (MS-STFT) discriminator with $5$ different scales~\cite{encodec}.
The FFT sizes and window lengths are set to $[4096, 2048, 1024, 512, 256]$, and their hop lengths are given as $[1024, 512, 256, 128, 64]$. 
We follow \cite{diffsep} in applying the diffusion process in the time domain, instead of the non-linear spectrogram domain used in \cite{sgmse, se_sb, se_sb_sym, sgmse_sb}, so our discriminator operates on white Gaussian noise.

For denoising, we use the same reconstruction loss as in \cite{se_sb, sgmse_sb, sbse_workshop}, given by \( L_\mathrm{recon} = ||\mathbf{X}_0' - \mathbf{X}_0||_2^2 + \alpha ||\mathbf{x}_0' - \mathbf{x}_0||_1 \), where \( \textbf{X}_0' \) is the estimated power-compressed complex spectrogram output by the generator, and \( \mathbf{x}_0' \) is the corresponding time-domain signal obtained through inverse STFT. \( \mathbf{X}_0 \) and \( \mathbf{x}_0 \) are the corresponding targets. 
For dereverberation, since the model does not train effectively with simple losses alone, we additionally incorporate the multi-scale mel loss, following the setup in \cite{dac}. 
The weighting coefficients for reconstruction loss are set as follows: $\alpha = 10^{-3}$, $0.01$ for the additional multi-scale mel loss for dereverberation, and $\lambda_\mathrm{recon} = 100$.

\section{Evaluation}
\subsection{Experimental Setup}
We trained our model with a batch size of 8, and the optimizer was AdamW~\cite{adamw} with a learning rate of $10^{-4}$. We use the exponential moving average (EMA) of the weights with a decay of 0.999~\cite{diffusion_ema}. The average SI-SDR of 20 randomly selected validation examples is logged during training, and the best-performing model is chosen for evaluation. All models are trained on 4 NVIDIA RTX 4090 or 2 A6000 GPUs. 

\subsection{Denoising Results}
We trained and tested the models on the full-band EARS-WHAM dataset~\cite{ears} for the denoising task. We used the ODE sampler as it has demonstrated better performance for the denoising task~\cite{se_sb}. Results were evaluated using intrusive metrics: Perceptual Evaluation of Speech Quality (PESQ)~\cite{pesq}, Scale-Invariant Signal-to-Distortion Ratio (SI-SDR)~\cite{sisdr}, and Extended Short-Time Objective Intelligibility (ESTOI)~\cite{estoi}. Additionally, we report non-intrusive metrics: DNSMOS~\cite{dnsmos} and SIGMOS~\cite{sigmos}, where SIGMOS is included because it supports full-band signals, unlike DNSMOS.
We compared our proposed model with the following baselines: Conv-TasNet~\cite{convtasnset} and Demucs~\cite{demucs} for the predictive models and CDiffuSE~\cite{cdiffuse}, SGMSE+~\cite{sgmse}, and SB~\cite{sbse_workshop} for the generative models.

\begin{table}[t]
\setlength\tabcolsep{4pt}
\centering
\caption{Denoising results on EARS-WHAM.}
\vspace{-0.3cm}
\label{tab:results}
\resizebox{1.0\linewidth}{!}{%
\begin{tabular}{lc|cccccc}
\toprule
\textbf{Model} & \textbf{NFEs} & \textbf{PESQ} & \textbf{SISDR} & \textbf{ESTOI} & \textbf{DNSMOS} & \textbf{SIGMOS} \\ 
\midrule 
Unprocessed & & 1.23 & 5.88 & 0.49 & 2.76 & 1.94 \\
\midrule
Conv-TasNet~\cite{convtasnset} & 1 & 2.31 & 16.9 & 0.70 & 3.47 & 2.69 \\
CDiffuSE~\cite{cdiffuse} & 6 & 1.60 & 8.35 & 0.53 & 2.87 & 2.08 \\
Demucs~\cite{demucs} & 1 & 2.37 & 16.9 & 0.71 & 3.66 & 2.87 \\
\midrule 
SGMSE+~\cite{sgmse} & 60 & 2.49 & 16.8 & 0.73 & \textbf{3.91} & 3.42 \\
\midrule
SB~\cite{sbse_workshop} & 50 & 2.32 & \textbf{17.9} & 0.73 & 3.87 & 3.44 \\
 & 4 & 2.32 & 16.3 & 0.72 & 3.84 & 3.34 \\
 & 1 & 2.20 & 13.4 & 0.67 & 3.68 & 3.15 \\
\midrule
SB-UFOGen & 4 & 2.52 & 17.7 & \textbf{0.74} & 3.88 & \textbf{3.49} \\
 & 1 & \textbf{2.56} & \textbf{17.9} & \textbf{0.74} & 3.88 & 3.48 \\ 
\bottomrule
\end{tabular}%
}
\vspace{-0.6cm}
\end{table}

The results on the EARS-WHAM test set are presented in Table~\ref{tab:results}. Our model, SB-UFOGen, significantly outperforms competing models across all evaluation metrics when using the same sampling steps, 4 or 1. Moreover, SB-UFOGen, with only a single step, even outperforms the baselines that require 50 steps or more to achieve their best performance. 
We also calculate the Real-Time Factor (RTF) by measuring the processing time per second, averaged over 20 test set utterances, using an NVIDIA RTX A6000 GPU. While the SB model~\cite{sbse_workshop} with 50 iterations achieves an RTF of $2.112$ proc/s, our SB-UFOGen model with a single iteration achieves $0.046$ proc/s, making it competitive with predictive models.

\begin{figure}[t]
    \centering
    \includegraphics[width=.95\linewidth]{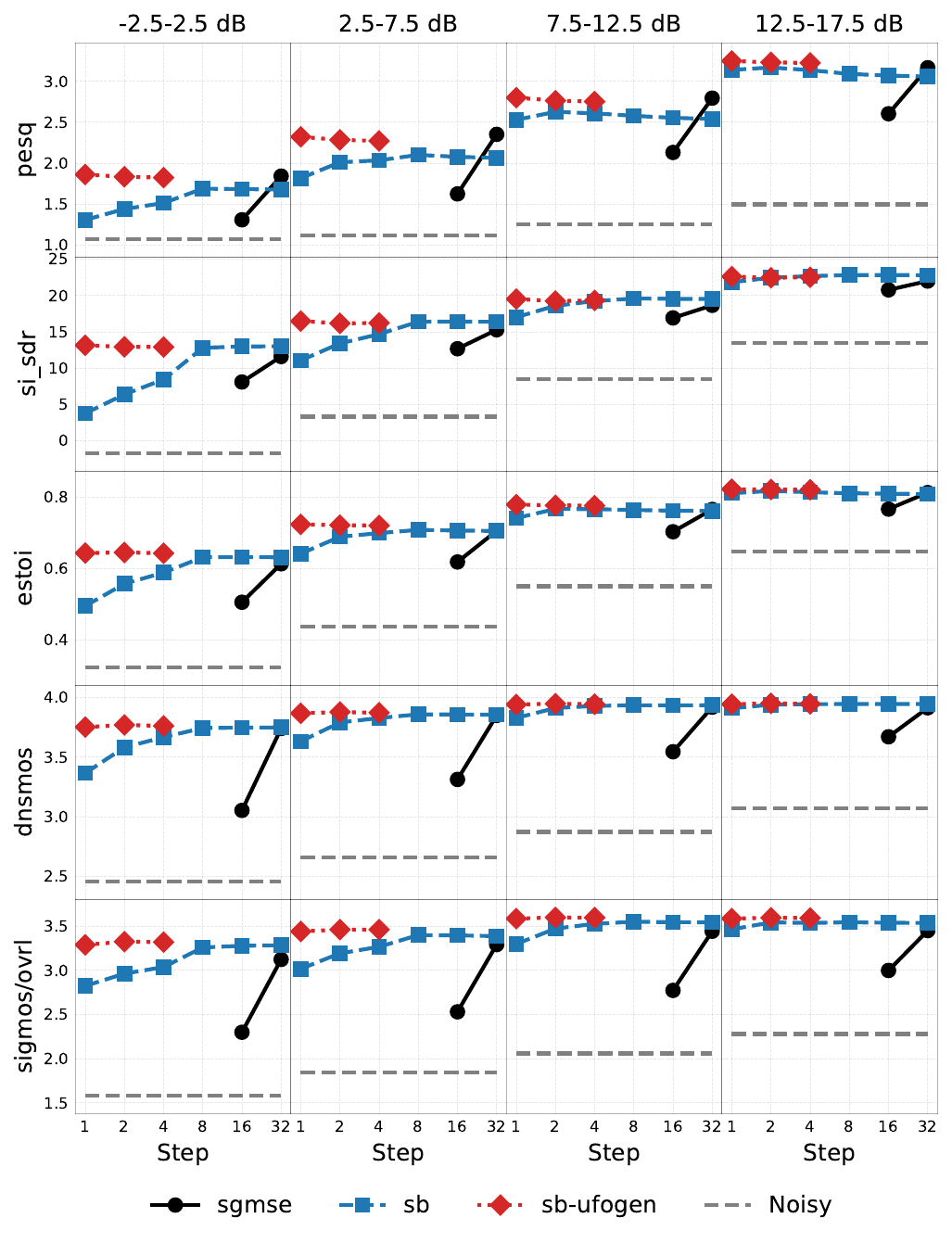}
    \vspace{-0.45cm}
    \caption{Results on EARS-WHAM test set across different sampling steps and SNR ranges.}
    \vspace{-0.29cm}
  \label{fig:metrics_plot_graph}
\end{figure}

Figure~\ref{fig:metrics_plot_graph} shows the influence of the number of steps in the reverse process on the EARS-WHAM test dataset, segmented by input SNR. While SGMSE+ suffers significant performance degradation as the number of steps decreases, the SB model with the ODE sampler effectively maintains denoising performance with fewer iterations, especially in high-SNR scenarios. However, for low SNR, the performance of SB still declines sharply with fewer iterations. Comparing 32-step and 1-step inference of SB, SI-SDR decreases by only $0.97$ dB and SIGMOS by $0.07$ in the SNR range of $[12.5, 17.5]$ dB, whereas in $[-2.5, 2.5]$ dB range, the decreases are drastic, with a drop of $9.23$ dB and $0.46$, respectively. 
Therefore, diffusion or SB-based speech enhancement models still require increasing the number of sampling steps, which is not the case for our model.

\begin{table}[t]
\setlength\tabcolsep{4pt}
\centering
\caption{Denoising results on EARS-WHAM (with PESQ loss).}
\vspace{-0.3cm}
\label{tab:results_pesq}
\resizebox{0.98\linewidth}{!}{%
\begin{tabular}{lc|ccccc}
\toprule
\textbf{Model} & \textbf{NFEs} & \textbf{PESQ} & \textbf{SISDR} & \textbf{ESTOI} & \textbf{DNSMOS} & \textbf{SIGMOS} \\ 
\midrule
SB~\cite{se_sb} & 50 & 3.08 & 16.2 & 0.73 & 3.74 & \textbf{3.17} \\
 & 4 & 3.03 & 15.9 & 0.72 & 3.72 & 3.10 \\
 & 1 & 3.09 & 16.1 & 0.73 & 3.71 & 3.10 \\
 \midrule
SB-UFOGen & 4 & 3.13 & 17.1 & \textbf{0.74} & \textbf{3.76} & 3.16 \\
 & 1 & \textbf{3.14} & \textbf{17.2} & \textbf{0.74} & \textbf{3.76} & 3.15 \\ 
 \bottomrule
\end{tabular}%
}
\vspace{-0.55cm}
\end{table}

Since PESQ is regarded as a standard evaluation metric for speech enhancement, a differentiable version, PESQ loss \cite{torchpesq}, 
was introduced in~\cite{sgmse_sb, sbse_workshop}.
Given its state-of-the-art performance on the PESQ metric, we also evaluate another version of our model with PESQ loss, using the same weighting coefficient as in \cite{sbse_workshop}.
The results are presented in Table \ref{tab:results_pesq}. Unlike the formal results, the SB model is more robust to the number of sampling steps when incorporating the PESQ loss. Nevertheless, the SB-UFOGen model outperforms the SB model.
On the other hand, incorporating the PESQ loss leads to a decline in other metrics, particularly non-intrusive metrics such as DNSMOS and SIGMOS. This is expected, as PESQ loss only prioritizes PESQ, increasing the risk of overfitting to this specific metric~\cite{pesqetarian}.

\subsection{Dereverberation Results}
We also trained and tested the models using the full-band EARS-Reverb dataset~\cite{ears} for the dereverberation task. We use the SDE sampler as it has demonstrated superior performance for the dereverberation task~\cite{se_sb}. Results are evaluated using the same metrics as for denoising, except that SIGMOS Reverb is reported instead of the DNSMOS.

\begin{table}[t]
\setlength\tabcolsep{4pt}
\centering
\caption{Dereverberation results on EARS-Reverb.}
\vspace{-0.3cm}
\label{tab:results_reverb}
\resizebox{0.98\linewidth}{!}{%
\begin{tabular}{lc|cccccc}
\toprule
\textbf{Model} & \textbf{NFEs} & \textbf{PESQ} & \textbf{SISDR} & \textbf{ESTOI} & \textbf{SIGMOS} & \textbf{MOS Reverb} \\ 
\midrule 
Unprocessed & & 1.47 & -16.22 & 0.52 & 2.77 & 2.97 \\
\midrule 
SGMSE+~\cite{sgmse} & 60 & 3.04 & 6.35 & 0.85 & \textbf{3.49} & \textbf{4.73} \\
\midrule
SB~\cite{se_sb} 
 & 50 & \textbf{3.41} & 6.65 & \textbf{0.88} & 3.37 & \textbf{4.73} \\
 & 4 & 3.24 & 0.94 & 0.86 & 3.32 & 4.71 \\
 & 1 & 2.12 & -12.31 & 0.77 & 3.16 & 4.47 \\
\midrule
SB-UFOGen 
 & 4 & 3.33 & 6.71 & 0.87 & 3.37 & 4.72 \\
 & 1 & 3.36 & \textbf{8.73} & \textbf{0.88} & 3.33 & 4.71 \\ 
\bottomrule
\end{tabular}%
}
\vspace{-0.2cm}
\end{table}

\begin{figure}[t]
    \centering
    \vspace{-0.25cm}
    \includegraphics[width=\linewidth]{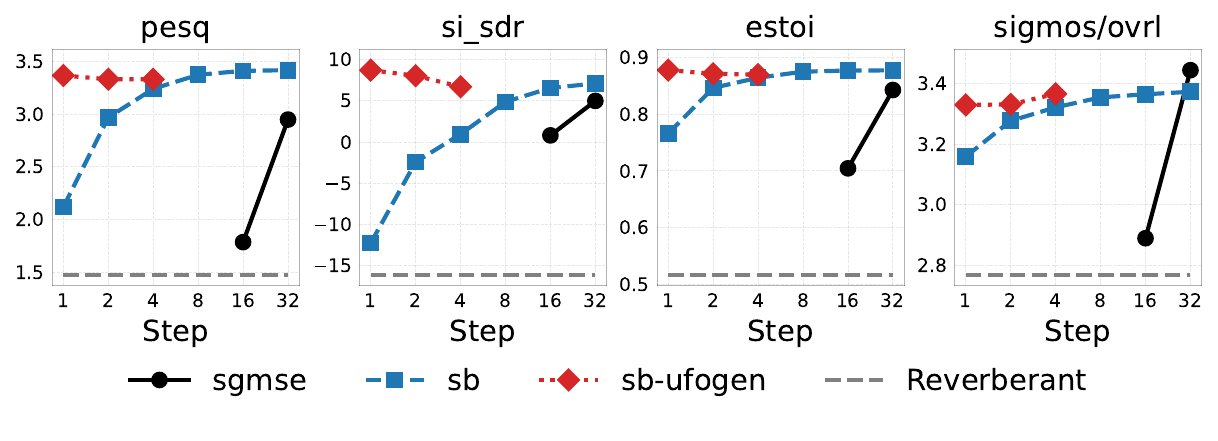}
    \vspace{-0.85cm}
    \caption{EARS-Reverb results across different sampling steps. }
    \vspace{-0.6cm}
  \label{fig:metrics_plot_graph_reverb}
\end{figure}

\looseness=-1
The results on the EARS-Reverb test set are presented in Table \ref{tab:results_reverb}. SB-UFOGen also significantly outperforms baselines on the dereverberation task when using the same sampling steps, 4 or 1, especially in terms of SI-SDR.  Moreover, single-step SB-UFOGen achieves performance comparable to baselines that require 50 steps or more to reach their optimal results.

Figure~\ref{fig:metrics_plot_graph_reverb} illustrates the impact of the number of steps in the reverse process on the EARS-Reverb test dataset. While the SGMSE+ model experiences significant performance degradation as the number of steps decreases, the SB model shows some robustness to the number of steps. However, the SB model yet shows a decline in performance with fewer iterations. Therefore, for dereverberation tasks based on the SB model, reducing the number of sampling steps remains a challenge, which our model effectively addresses.

\section{Conclusion}
Diffusion models have gained attention in speech enhancement for their generative capabilities, while SB models have recently emerged as a promising alternative. However, both approaches are still constrained by the iterative reverse process. Our investigation reveals that SB models suffer notable performance decline, particularly under low-SNR conditions. 
To address this, we combine SB and GAN models, achieving better performance than the baselines with single-step inference for both denoising and dereverberation.
Notably, single-step inference, as a direct predictive mapping, yields higher scores on intrusive metrics than 4-step inference.
In future work, similar studies can be conducted under more severe conditions and a wider range of degradations. Moreover, our model can be extended to handle a broader range of degradations, including super-resolution~\cite{nuwave} and universal speech enhancement~\cite{universe}.

\section{Acknowledgements}
This work was supported by the National Research Foundation of Korea (NRF) grant funded by the Korea government (MSIT) (No. RS-2024-00461617, 50\%), and the Institute of Information \& Communications Technology Planning \& Evaluation (IITP) grant funded by the Korea government (MSIT) [No. RS-2022-II220641, 40\%], [No. RS-2021-II211343, Artificial Intelligence Graduate School Program (Seoul National University), 5\%], and [No. RS-2021-II212068, Artificial Intelligence Innovation Hub, 5\%].

\bibliographystyle{IEEEtran}
\bibliography{mybib_abbv}

\end{document}